# Cooperative Multi-spacecraft Observation of Incoming Space Threats

**Ravi teja Nallapu and Jekanthan Thangavelautham**

*Space and Terrestrial Robotic Exploration (SpaceTREx) Laboratory, University of Arizona*

**ABSTRACT**

Earth is constantly being bombarded with material from space. Most of the natural material end up being dust grains that litter the surface of Earth, but larger bodies are known to impact every few decades. The most recent large impact was Chelyabinsk which set off a 500-kiloton explosion which was 40 times that of the Hiroshima nuclear explosion. Apart from meteors, there is a growing threat of space assets deorbiting. With these impending space threats, it is critical to have a constellation of satellites to autonomously lookout for meteors and reentering space debris. By using multiple spacecraft, it is possible to perform multipoint observation of the event. Through multipoint observation, it is possible to triangulate the location of the observed event. The detection, tracking, and analysis of these objects all need to be performed autonomously. Our previous work focused on developing several vision algorithms including blob-detection, feature detection, and neural network-based image segment classification. For this multipoint observation to occur, it requires multiple spacecraft to coordinate their actions particularly fixating on the space observation target. Furthermore, communication and coordination are needed for bringing new satellites into observation view and removing other satellites that have lost their view. In this paper, we analyze state-of-the-art observation technology for small satellites and perform detailed design of its implementation. Through this study, we estimate the error estimates on position, velocity, and acceleration. We presume use of low to mid-tier cameras for the spacecraft. We then analyze the implications of multiple spacecraft and see how the estimates will be improved with enough crafts. With a critical number of spacecraft, we hope to place a bound on the errors and then determine what else can be done to impact the overall capabilities. In this study, we first present a new meteor localization algorithm which can be deployed on a spacecraft constellation. We then present a sensitivity study of the developed algorithm to constellation parameters. Following this, we present an automated architecture to design optimal meteor monitoring constellations under real-life constraints such as spacecraft outages and detection requirements. We also present a dynamic simulation architecture using the STK-MATLAB interface. Finally, we conclude by identifying pathways forward to advance the algorithms discussed to improve space situational awareness.

## 1. INTRODUCTION

Meteor showers are common events that occur throughout the year. These are caused when meteoroids, the ejected fragments of asteroids or comets, enter the Earth's atmosphere. Most of these fragments are small and end up as dust grains on the surface of Earth, and therefore are harmless. However, the larger bodies that release massive amounts of energy upon their entry are not uncommon. The database of near-Earth objects maintained by NASA-JPL, reports that at least 600 meteor events with energies greater than 0.1 kilotons were recorded in the last 30 years [1]. The atmospheric explosion of the Chelyabinsk meteor in 2013, where about 500 kilotons of energy were released during its airburst [2], serves as an indicator of the potential hazard of these impacts. The fallout of such impact events can be catastrophic. A similar problem also exists from the reentering space debris [3]. Therefore, there is a strong need to have a real-time monitoring network to monitor incoming meteor events.

Modern-day meteor studies are conducted from a network of ground-based observatories including radar. These ground observatories can triangulate the meteor event and can also partially study the composition of the meteor. However, these observations are limited by the Field-of-View (FoV) of the individual instruments, and Earth's atmospheric interference [4]. These challenges can be overcome by deploying a constellation of orbiting spacecraft. However, state-of-the-art event localization algorithms are still deployed from a stationary ground observatory reference and must be developed from a dynamic orbiting reference frame, which is a non-trivial task. Then, there is the problem of designing an optimal constellation which would achieve the required probability of detection when subjected to real-world constraints such as spacecraft outages [5]. This work addresses the two challenges described above. We begin by describing our approach to the localization problem when an entry event is detected by two or more spacecraft. Following this, we validate the algorithm using Satellite Tool Kit (STK) simulations.

We then proceed to the optimal constellation design problem, where we present an automated design architecture for designing meteor monitoring constellations which meet a required probability of success when subjected to real-world constraints such as spacecraft outages. The organization of the paper is as follows: Section 2 presents an overview of related work carried out in developing meteor monitoring constellations. Section 3 presents the methodology used in the current work. Here we present the meteor localization algorithm which would be deployed onboard the orbiting spacecraft. We then proceed to a case study where the algorithm is applied to a simulated entry event. Following this, we present the architecture of an automated constellation design algorithm. Section 4 describes the results of the localization study and then apply the lessons learned to design a meteor monitoring constellation. Finally, Section 5 will discuss the contributions of the current work and future work required in design constellations to improve space situational awareness.

## 2. RELATED WORK

Most meteor showers have been known to be caused by objects whose diameter is greater than 2 mm. These events have been observed at altitudes ranging from $70 - 140$ km and where the entry velocities range from $11 - 72$ km/s. Photometric observations of these meteors have been shown to provide plenty of insight into the entering body. If a single camera is used, the observations can allow us to estimate the visual magnitude of the meteor. In the case of meteor showers, these observations can measure the entry rate and population index. However, with two or more spacecraft observing the same event, the visual depth of the event can be recovered [4]. These multi-point observations allow for triangulating the location of the event by fitting the best possible plane in the direction of the observations [6]. Furthermore, the material composition, strength properties and cohesion of meteor are expected to be analyzed in upcoming missions using on-orbit centrifuge laboratories [26-27].

The entry velocity can be estimated by generating a time history of the event location. Once the state of the event at a time frame is known, the dynamics of the body can be propagated forward and backward in time to estimate the trajectory of the meteor. Backward propagation can help us identify the source of the meteor, particularly its heliocentric trajectory. This can even help us identify the parent source of the meteor, while the forward propagation can help us identify the dark flight of the meteor and, as a result, help identify the location of residual meteorites (if any). Event localization would also provide an insight into statistical information such as photometric mass, length, and diameter of the event in real-time [4]. Most meteor observations to date have been conducted through networks of ground-based observatories that are distributed all around the world [7, 8, 9]. However, ground-based observations are limited due to the limited field of view (FoV) of the ground site instruments. However, this limitation can be overcome if the network of observatories were deployed as a constellation of multiple orbiting spacecraft.

Spacecraft constellations have been well studied in the literature for applications requiring spatial and temporal coverage. Many satellite constellations have been realized for navigation [10], communication [11], and weather monitoring applications [12]. Several constellation architectures that are based on the distribution of the participating spacecraft have also been proposed [13, 14]. The most popular constellation architecture is the Walker-Delta constellation, where groups of spacecraft are placed in circular orbits which are uniformly distributed in their right ascension of the ascending node (RAAN), true anomaly separations, and adjacent plane spacing [15]. A Walker-Delta constellation contains $N_p$ circular orbits, where each orbit has $N_{sp}$ spacecraft. Therefore, the total number of spacecraft in the constellation is:

$$T = N_p N_{sp} \qquad (1)$$

All the orbits will have the same inclination $i_n$. Additionally, the Walker constellations have a spacing parameter to avoid collisions between spacecraft in different planes given by:

$$\Delta\phi = F \frac{360}{T} \qquad (2)$$

Where $F$ is a positive integer. Thus, the design of a Walker constellation is specified in the format $in: T/N_p/F$. The geometry of an example Walker constellation showing the different design parameters is presented in Fig 1. The semi-major axis for deploying these constellations is typically selected to achieve repeat ground track (RGT) orbits which will result in periodically repeating the motion of the participating spacecraft [16].

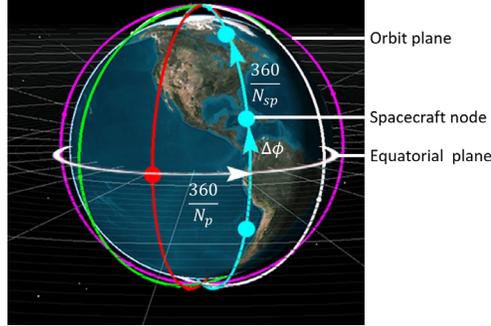

Fig. 1. The geometry of a Walker-Delta constellation of the pattern 90: $T/N_p/F$ showing the significance of different design parameters.

Specifically, the spacecraft will complete $N_{orb}$ orbits for every $N_{rot}$ rotations of the Earth. Current state-of-the-art algorithms include RGT orbits in the presence of orbital perturbations such as oblateness [17]. Our previous work focused on several aspects of deploying a meteor monitoring constellation. We proposed the SWIMSat mission, which is a CubeSat with a wide FoV imaging camera as a prototype meteor-tracking demonstrator spacecraft of the constellation [18]. We developed two detection algorithms to identify and track a meteor event, which used an edge detection technique [19] and a color-based thresholding technique [20]. The algorithms were tested in the laboratory under controlled conditions using a hardware testbed [21]. In [22], we examined the possibilities of designing a SWIMSat constellation at different RGT orbits using a grid search optimization scheme. The current work will extend our previous work by developing algorithms which will be deployed on the SWIMSat constellation to localize the meteor event, and then use improved optimization schemes to design robust monitoring constellations.

## 3. METHODOLOGY

This section describes the methodology used in the current work to design meteor monitoring constellation. We begin by providing a brief derivation of the localization algorithm, and then describe a validation study to test these algorithms. We then present the constellation design as an optimization problem when subjected to real-world constraints such as meeting the required detection probability, robustness to spacecraft outages.

**Multi-spacecraft event localization:** Let a meteor event $M$ be observed by $N_{sc}$ spacecraft located at the cartesian coordinates $(x_j, y_j, z_j)$ in an inertial frame. We assume that the spacecraft cameras are calibrated, which means that each pixel inside the FoV of the camera corresponds to right ascension and declination angles measured in an inertial reference frame. Now, considering that the observed meteor spans a set of $N_{px,j}$ pixels on each camera which correspond to right ascension angles $\alpha_{i,j}$ and declination angles $\delta_{i,j}$. The $j$ is an indicator of the imaging spacecraft while index $i$ is an indicator of the pixels in each camera, i.e. $j = 1, 2, \ldots, N_{sc}$ and $i = 1, 2, \ldots, N_{px,j}$. We can now define a unit vector in the direction of $(\alpha_{i,j}, \delta_{i,j})$ through its cartesian components

$$\begin{aligned} \xi_{i,j} &= \cos \delta_{i,j} \cos \alpha_{i,j} \\ \eta_{i,j} &= \cos \delta_{i,j} \sin \alpha_{i,j} \\ \zeta_{i,j} &= \sin \delta_{i,j} \end{aligned} \quad (3)$$

The FoV of a spacecraft camera upon detecting a meteor event and its corresponding observations are shown in Fig 2.

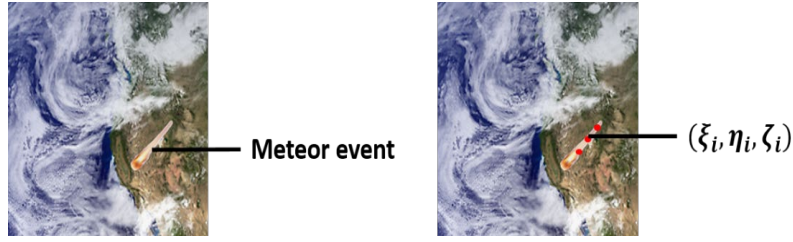

Fig. 2. Field of view of a monitoring spacecraft camera (left) and its direction measurements from the observed pixels (right).

The meteor event will be estimated to lie in a plane perpendicular to the Line of Sight (LoS) from the spacecraft to one of the meteor pixels such that:

$$a_j \xi_{i,j} + b_j \eta_{i,j} + c_j \zeta_{i,j} = \Delta_{i,j} \quad (4)$$

Where $a_j$, $b_j$, $c_j$, and $\Delta_{i,j}$ are parameters that define a plane. Ceplecha et al. [6] provide a closed-form expressions for the coefficients $a_j$, $b_j$, $c_j$ which minimize the magnitude of $\Delta_{i,j}$ as follows:

$$a_j' = \sum_{i=1}^{N_{px,j}} (\xi_{i,j}\eta_{i,j}) \sum_{i=1}^{N_{px,j}} (\zeta_{i,j}\eta_{i,j}) - \sum_{i=1}^{N_{px,j}} (\eta_{i,j}^2) \sum_{i=1}^{N_{px,j}} (\xi_{i,j}\zeta_{i,j})$$

$$b_j' = \sum_{i=1}^{N_{px,j}} (\xi_{i,j}\eta_{i,j}) \sum_{i=1}^{N_{px,j}} (\zeta_{i,j}\xi_{i,j}) - \sum_{i=1}^{N_{px,j}} (\xi_{i,j}^2) \sum_{i=1}^{N_{px,j}} (\eta_{i,j}\zeta_{i,j})$$

$$c_j' = \sum_{i=1}^{N_{px,j}} (\xi_{i,j}^2) \sum_{i=1}^{N_{px,j}} (\eta_{i,j}^2) - \left(\sum_{i=1}^{N_{px,j}} (\xi_{i,j}\eta_{i,j})\right)^2$$

$$d_j' = \sqrt{a_j'^2 + b_j'^2 + c_j'^2} \quad (5)$$

$$a_j = \frac{a_j'}{d_j'}$$

$$b_j = \frac{b_j'}{d_j'}$$

$$c_j = \frac{c'}{d_j'}$$

The coefficients $a_j$, $b_j$, $c_j$, can be packed as the vector $\hat{n}_j = [a_j \; b_j \; c_j]^T$ which will represent the normal vector to the computed plane. Additionally, the offset parameter of the plane can be computed as:

$$d_j = -(a_j x_j + b_j y_j + c_j z_j) \quad (6)$$

With the plane parameters, the next step would be to compute the direction of the meteor or its radiant. The radiant is a unit vector along the length of the pixels, as seen in the viewing plane of the spacecraft cameras. Since any two non-parallel planes intersect along a line, the radiant can be defined through the intersection of any two planes computed above. This presents us with $\binom{N_{sc}}{2}$ possible combinations, which can be chosen based on any user-defined selection scheme. In this work, we select the two planes that have the least $\Delta_{i,j}$ corresponding to its measurements. Let the two spacecraft be referred to as spacecraft $A$ and spacecraft $B$ located at cartesian coordinates $(x_A, y_A, z_A)$ and $(x_B, y_B, z_B)$ respectively. The radiant can now be defined as the intersection of the planes in (5) for the spacecraft $A$ and $B$ as shown in Fig 3.

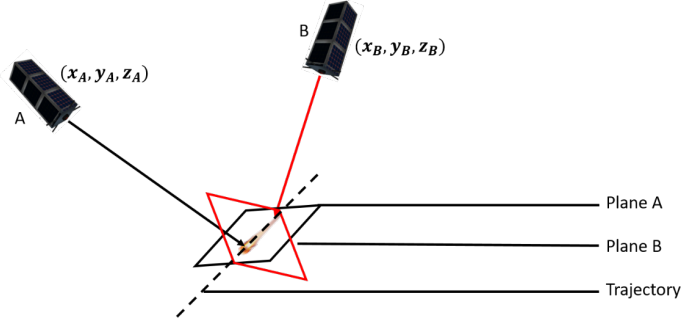

Fig. 3. Computation of the meteor trajectory or the radiant from the observations of two spacecraft.

To compute the radiant, we define normal vectors to the planes computed through spacecraft $A$ and $B$ as $\hat{n}_A = [a_A \; b_A \; c_A]^T$ and $\hat{n}_B = [a_B \; b_B \; c_B]^T$ respectively, where the elements of the normal vector are computed from (5) and correspond to their respective spacecraft. The radiant vector is now given by:

$$\widehat{M}_R = \frac{\hat{n}_A \times \hat{n}_B}{|\hat{n}_A \times \hat{n}_B|} \quad (7)$$

A key advantage of this algorithm is that the spacecraft need not look at the exact same point in space. Once the radiant is computed, the final step of the localization problem is to determine the physical locations of the meteor points which span the camera pixels of the spacecraft. The physical location of the points can be obtained by projecting the LoS vector to a specific pixel along the radiant line in a plane perpendicular to the LoS as shown in Fig 4.

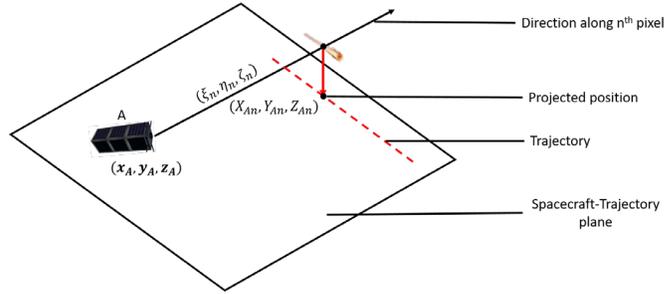

Fig. 4. Geometric interpretation of the localization algorithm where the pixel direction is projected onto the radiant line.

As seen in Fig. 4, the projected plane contains the location of the spacecraft, the radiant line, and the projected position of the pixel. Let us assume that we are interested in determining the location of the $n^{th}$ pixel located in the FoV of spacecraft $A$. Then the equation of the plane can be expressed as:

$$a_n x_n + b_n y_n + c_n z_n + d_n = 0 \quad (8)$$

Where,

$$\begin{aligned}
a_n &= \eta_n c_A - \zeta_n b_A \\
b_n &= \zeta_n a_A - \xi_n c_A \\
c_n &= \xi_n b_A - \eta_n a_A \\
d_n &= -(a_n X_A + b_n Y_A + c_n Z_A)
\end{aligned}$$

Where $(\xi_n, \eta_n, \zeta_n)$ are the measurements of pixel $n$ in the FoV of camera $A$ through (3). Since the projection of the point corresponding to the physical location of pixel $n$ lies on the plane described in (8), and all the planes computed in (5) and (6), the best estimate of the location of pixel $n$ in the inertial frame can be obtained by solving the overdetermined linear system

$$\begin{bmatrix} a_1 & b_1 & c_1 \\ \vdots & \vdots & \vdots \\ a_{N_{sc}} & b_{N_{sc}} & c_{N_{sc}} \\ a_n & b_n & c_n \end{bmatrix} \begin{bmatrix} x_n \\ y_n \\ z_n \end{bmatrix} = - \begin{bmatrix} d_1 \\ \vdots \\ d_{N_{sc}} \\ d_n \end{bmatrix} \quad (9)$$

This is the geocentric location of a point on the meteor and can be estimated by solving the linear system in (9).

**Accuracy of localization:** The obvious question that any localization algorithm faces is: 'How accurate can is the estimated position?'. To answer this question, we develop a Monte-Carlo simulation architecture in MATLAB which simulates a random meteor detection, followed by the application of the algorithm described above to localize the event and then study the localization errors. We begin by noting that there are three critical parameters that can influence the localization accuracy. The number of spacecraft observing the event $N_{sc}$, the pointing accuracy of the spacecraft $P_{ac}$ which limits the ability to measure the right ascension and declination angles, and the accuracy in the knowledge of the locations of the spacecraft in the inertial space $(x_j, y_j, z_j)$. In this study, we will study the sensitivity of the localization accuracy to the spacecraft pointing accuracy and the number of spacecraft making the measurement. We achieve this by randomly simulating meteor entry events in the FoV of the spacecraft for a Monte Carlo simulation with $N_{MC,1}$ simulations, running the algorithm for different $N_{sc}$ and $P_{ac}$, and finally examining the results.

There are four critical components of the simulation: (a) locating the spacecraft, (b) generating the meteor, (c) computing the spacecraft measurements, and (d) obtaining the distribution of localization errors. The $N_{sc}$ spacecraft in the simulation are generated randomly at a fixed altitude $h_{sc}$ above the Earth's surface. To make valid observations of the event, the spacecraft are located relatively closely with respect to each other such that their latitudes and longitudes have bounded random differences. The head of the meteor is generated from the centroid of the imaging spacecraft, to ensure that a spacecraft has a direct LoS with respect to the meteor. The altitude of the head is uniformly distributed between 70 to 140 km above the Earth's surface. The meteor trail is generated by drawing a line of length $l_M$ in an arbitrary direction. We then generate $N_{pt}$ equally spaced points on this line which will be assumed to be imaged by the spacecraft camera. The measurement of right ascension and declination will be simulated by computing the true LoS vector with respect to each spacecraft and the meteor points. The true right ascension and declination are the spherical coordinate angles of the computed LoS vectors. The measurement will be simulated by adding a uniformly distributed pointing noise along to the truth. If we assume that $\alpha_{i,j}$ and $\delta_{i,j}$ are the true right ascension and declination angles, and $\bar{\alpha}_{i,j}$ and $\bar{\delta}_{i,j}$ are their respective measurements, then:

$$\begin{aligned} \bar{\alpha}_{i,j} &= \alpha_{i,j} + E(\alpha_{i,j}: -P_{ac}, P_{ac}) \\ \bar{\delta}_{i,j} &= \delta_{i,j} + E(\delta_{i,j}: -P_{ac}, P_{ac}) \end{aligned} \quad (10)$$

Where $E(x: l, u)$ denote a uniformly distributed error in the parameter $x$, bounded between $l$ and $u$. We estimate the location of all the meteor points using the algorithm described above. A simulated experiment with $N_{sc} = 5$ spacecraft localizing a meteor is shown in Fig. 5.

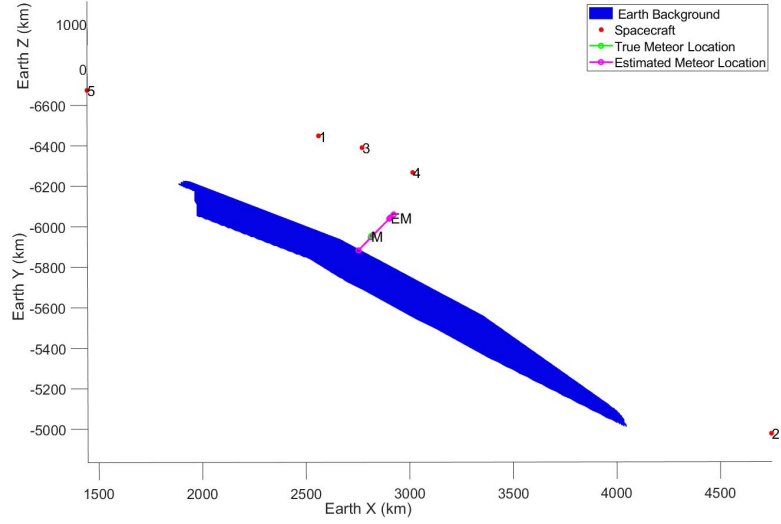

Fig. 5. A representative test case generated during the trade study where randomly generated meteor points from the set $M$ are observed by 5 spacecraft to produce the points on the estimated set $EM$.

Following this, we note the error as the magnitude in the difference of position vectors as follows:

$$e_i = |r_i - \bar{r}_i| \qquad (11)$$

Where, $r_i$ is the true position of the point $i$ whose corresponding measurement is $\bar{r}_i$. For each event, we record the maximum localization error, $\max(e_i)$, as the figure of merit and look at its distribution over $N_{MC,1}$ events for a given $N_{sc}$, and $P_{ac}$. This process is repeated over a grid spanned by the parameters given $N_{sc}$, and $P_{ac}$.

**Constellation Design:** The case study described above allows us to estimate the localization accuracy of $N_{sc}$ observing spacecraft when their pointing accuracy is limited to $P_{ac}$. The next step is to design a constellation which would ensure that a random meteor event will be imaged by at least $N_{sc}$ observing spacecraft. In this work, we develop an automated constellation design architecture to ensure the requirement. The constellation geometry will be similar to a Walker-Delta constellation but will allow elliptical orbits. The seed spacecraft will be assumed to be in a repeat ground track (RGT) orbit which is parameterized by a set of two positive integers $N_{orb}$ and $N_{rot}$. The integers will be used to obtain an initial guess of the spacecraft semi-major axis $a_{sc,0}$. The initial guess for the semi-major axis can be expressed using Kepler's third law as

$$a_{sc,0} = \left(\mu \left(\omega_E \left(\frac{N_{orb}}{N_{rot}}\right)\right)^2\right)^{\frac{1}{3}} \qquad (12)$$

Where, $\mu$ is the gravitational parameter of the Earth, and $\omega_E$ is the rotation rate of the Earth. The initial guess is then corrected for the Earth's oblateness effect using the algorithm described in Reference [17] to obtain the final semi-major axis $a_{sc}$ corresponding to an RGT orbit described by $N_{orb}$ and $N_{rot}$. The spacecraft will be in an eccentric orbit with eccentricity $e_{sc}$ at an inclination $i_{sc}$. The RAAN and true anomaly assignments are carried using the standard Walker-Delta conventions [15]. The argument of periapsis is set as 0 deg due to spherical symmetry involved. These six orbital elements are sufficient to describe the orbit of the seed spacecraft. The seed spacecraft is then used to define the Walker-Delta pattern $i_{sc}: N_p N_{sp}/N_p/F$. We assume that all spacecraft are nadir pointing and have a wide FoV enough to make observations with a minimum elevation angle $\varepsilon_{ob}$. The half FoV angle $\eta_{sc}$ of the spacecraft is then determined as:

$$\sin \eta_{sc} = \cos \varepsilon_{ob} \left(\frac{h_{sc}}{R_E + h_{sc}}\right) \qquad (13)$$

Where, $R_E$ is the radius of the Earth, and $h_{sc}$ is the spacecraft altitude. Since the altitude varies on an elliptical orbit, the periapsis altitude is used to compute the required FoV of the spacecraft, since it corresponds to the orbital location requiring the maximum FoV. While theoretically, the larger altitudes permit smaller FoVs, they also increase the aperture diameter $D_{sc}$ of the spacecraft camera for a specified resolution $x_r$. The constraint can be expressed as [16]:

$$D_{sc} = h_{max}\left(\frac{\lambda}{x_r}\right) \qquad (14)$$

Where, $h_{max}$ is the maximum or apoapsis altitude and $\lambda$ is the wavelength of the imaging spectrum. Since the wavelength of the visible spectrum is upper bounded by the red light, we use $\lambda = 0.7\ \mu m$ to estimate the worst-case aperture diameter required. Therefore, the design can be constrained by placing a requirement $D_{sc} \leq D_{max}$, where $D_{max}$ is specified by the user.

**Constellation coverage:** A meteor is randomly generated anywhere between the altitude ranges of 70 to 140 km on the surface of the Earth. The coverage to the meteor for all operating spacecraft is checked using the FoV clipping operation described in Reference 23. An event is said to be successfully detected if it falls inside the FoV of at least $N_{sc}$ operating spacecraft. The effectiveness of a constellation design is computed from a Monte Carlo simulation where $N_{MC,2}$ meteor events are randomly generated at meteoric altitudes and the detection by the constellation is verified. Let us assume that $N_D$ of these random events are detected by the constellation, we can define the effectiveness of the constellation as:

$$P_{suc} = \left(\frac{N_D}{N_{MC,2}}\right) \times 100 \qquad (15)$$

and where (15) allows us to define a quantitative basis to design constellations which are effective than a requirement $P_r$.

**Satellite outages:** For us to get realistic estimates of constellation effectiveness, we assume that during any random meteor event only $P_{op}$, a percentage of the satellites are operative, and the rest are defunct. Therefore only

$$N_{op} = \left(\frac{P_{op}}{100}\right) T = \left(\frac{P_{op}}{100}\right) N_p N_{sp} \qquad (16)$$

Spacecraft are used to detect a given event. The defunct satellites are randomly selected from a constellation design during each of the Monte Carlo runs.

**Optimal constellation design:** We can optimize the constellation by selecting the constellation parameters described above, such that it is able to meet the detection and observation criterion with a minimum number of spacecraft. This can be expressed as:

$$\min T = N_p N_{sp}$$

such that:

$$\begin{aligned} P_{suc}|_{P_{op}} &\geq P_r \\ D_{sc} &\leq D_{max} \end{aligned} \qquad (17)$$

Where $P_{suc}|_{P_{op}}$ specifies the effectiveness of a constellation, when at any time, only $P_{op}$ percentage of the satellites in the constellation are functioning. The constellation design variables in this process and their role is presented in a gene map format in Fig. 6.

| Parameter | # spacecraft orbits | # Earth days | Eccentricity | Inclination | # planes | # spacecraft per plane | Spacing parameter |
|---|---|---|---|---|---|---|---|
| Variable | $N_{orb}$ | $N_{rot}$ | $e_{sc}$ | $i_{sc}$ | $N_p$ | $N_{sp}$ | $F$ |
| Range | Integer $[1, N_{1,max}]$ | Integer $[1, N_{2,max}]$ | Real $[0, e_{max}]$ | Real $[0, \frac{\pi}{2}]$ | Integer $[1, N_{3,max}]$ | Integer $[1, N_{4,max}]$ | Integer $[1, N_{5,max}]$ |

The first two columns (# spacecraft orbits, # Earth days) form the **RGT orbit**. The first four columns belong to the **Seed spacecraft**, and the last three columns (# planes, # spacecraft per plane, Spacing parameter) form the **Walker constellation**.

Fig. 6. Gene map of the optimal constellation design problem showing different design variables and their significance.

We solve (17) using a mixed-integer genetic algorithm optimization solver in MATLAB. The bounds on design variables $N_{1,max}$, $N_{2,max}$, $N_{3,max}$, $N_{4,max}$, $N_{5,max}$ and $e_{max}$ are passed as user-defined parameters to the solver [24]. The obtained constellation is then passed to the Satellite Tool Kit (STK) software to perform further validation.

## 4. RESULTS & DISCUSSION

This section presents the results of the accuracy estimation simulations and followed by the design of an optimal constellation which meets a user-defined success criterion.

**Accuracy of localization:** The parameters used in the for studying the accuracy of the localization algorithm are presented in Table 1. The distribution of the mean and $1 - \sigma$ standard deviation of the maximum localization error over the input grid of observing spacecraft and pointing error of spacecraft camera is presented in Fig. 7.

Table 1: User-defined parameters passed to the localization error estimation simulations.

| Parameter | Value |
|---|---|
| Range of $N_{sc}$ grid | [2, 10] |
| Range of $P_{ac}$ grid | [1, 10] deg |
| Spacecraft altitude, $h_{sc}$ | 450 km |
| Length of meteor events, $l_M$ | 1 km |
| # meteor observation points, $N_{pts}$ | 6 |
| # Monte Carlo simulations, $N_{MC,1}$ | 10000 |

The distribution contours of the localization error in Fig. 7 presents some interesting insights for designing a meteor monitoring network in space. First, as expected the localization is large ($\geq 1,000$ km) when the spacecraft in the constellation have poor pointing performance. Additionally, just the bare minimum of using two spacecraft to localize an event also results in large errors. The accuracy of the estimates just starts to increase when three or more spacecraft are used to localize the event. The mean maximum error in case of a three spacecraft localization network can range from 50 km to 500 km depending on the pointing accuracy of the spacecraft. This can be seen from the sensitivity of the standard deviation contours, as the $1 - \sigma$ standard deviation drops from about 500 km when using two spacecraft to about 100 km when using three spacecraft to localize the event.

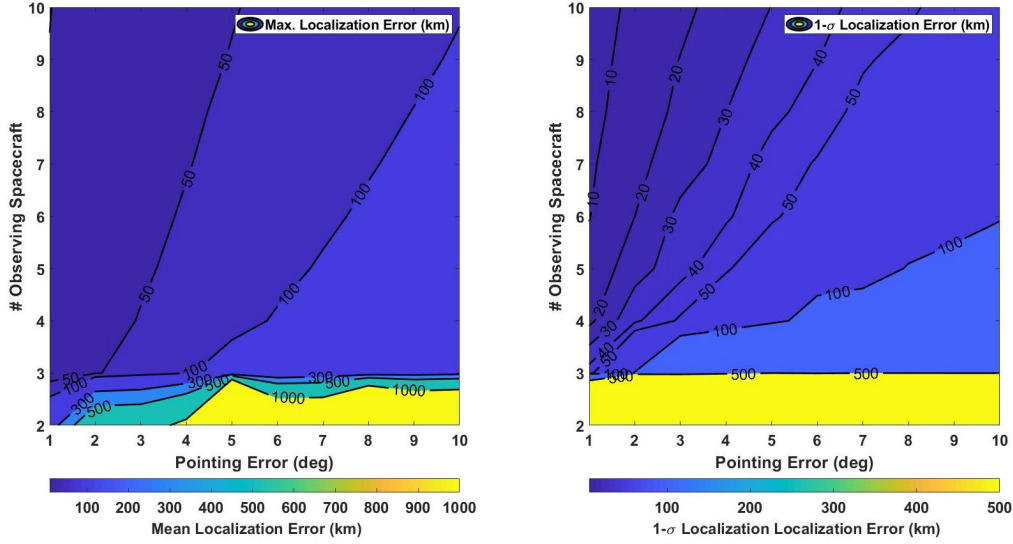

Fig. 7. Distribution of the mean maximum localization error (left), and its standard deviation (right) generated over the input grid of pointing error and observing spacecraft.

**Automated constellation design:** We now proceed to design an optimal constellation which can localize a random meteor event using at least three spacecraft. The constellation is expected to meet a minimum detection effectiveness requirement of 85 % when at any given time only 90 % of its total satellites are operative. The user-defined input parameters used for designing the optimal constellation are presented in Table 2. As mentioned earlier, the optimization problem is solved using the mixed integer genetic algorithm solver in MATLAB. Since the genetic algorithm solver is a stochastic optimizer, the problem is solved multiple times to verify the convergence of the results. Each optimizer run converged to an optimal design in 50 generations, where each generation spanned 100 individual designs. The best fitness in the evolution did not change for about 30 generations indicating convergence to an optimal design. The results of five optimizer runs showing the evolution of the mean and best design across different generations, along with a selected optimal gene is presented in Fig. 8.

Table 2: User-defined parameters passed as inputs to the optimal constellation design problem.

| Parameter | Value |
|---|---|
| Required detection efficiency, $P_r$ | 85 % |
| Operational percentage, $P_{op}$ | 90 % |
| Desired imaging resolution, $x_r$ | 1 m/pixel |
| Minimum elevation angle, $\varepsilon_{ob}$ | 5 deg |
| Maximum camera aperture diameter, $D_{max}$ | 1 m |
| Maximum spacecraft orbits, $N_{1,max}$ | 20 |
| Maximum Earth days, $N_{2,max}$ | 20 |
| Maximum orbit planes, $N_{3,max}$ | 20 |
| Maximum spacecraft per planes, $N_{4,max}$ | 20 |
| Maximum spacing parameter, $N_{5,max}$ | 5 |
| Maximum spacecraft eccentricity, $e_{max}$ | 0.8 |
| # Monte Carlo simulations, $N_{MC,2}$ | 10000 |

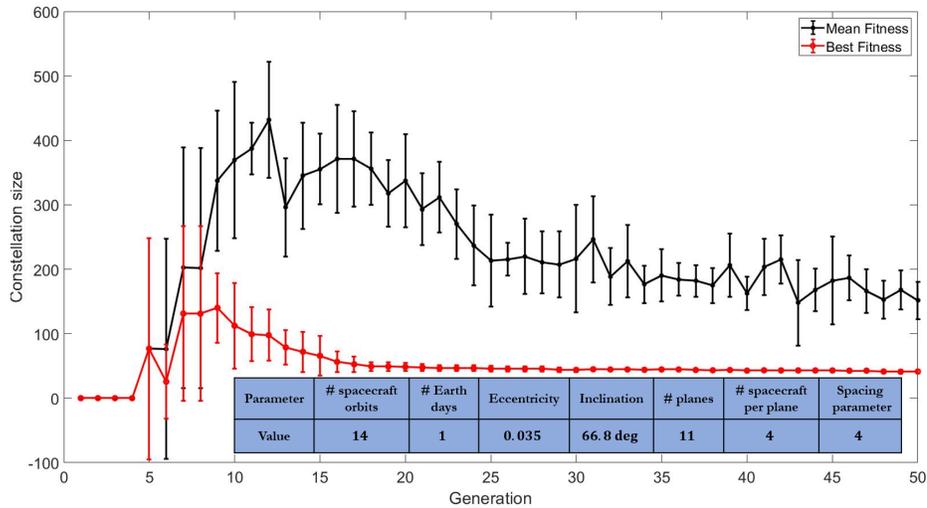

Fig. 8. Results of the optimizer runs showing the convergence of the optimal design, along with the selected optimal constellation gene.

As seen in Fig. 8, the selected optimal constellation contains a total of 44 spacecraft. The final selected constellation gene contained a 14: 1 RGT seed orbit with an eccentricity 0.035 which is inclined at 66.8 deg. The seed orbit was used to create a 66.8: 44/11/4 Walker-Delta constellation pattern. The selected constellation had detection effectiveness of 85.7 %, where observations were made with a maximum camera aperture diameter of 0.7 m thus meeting the design requirements. Due to the operational percentage, at any given time, there were 5 defunct spacecraft in the constellation. In one of the test cases of the constellation simulations, where a random meteor entry event was able to meet, $t$, the successful detection criteria of the constellation and is presented in Fig. 9.

**Design validation:** For us to verify the validity of the optimal constellation in dynamic conditions, the optimal constellation generated above was tested in STK. A simulation architecture was developed which was able to automate the validation process by passing the optimization parameters noted above to STK. The optimal gene parameters were used to create the Walker-Delta constellation. The meteor events were created as missile objects which were randomly initialized at altitudes between 70 to 140 km and on their ballistic downward flights.

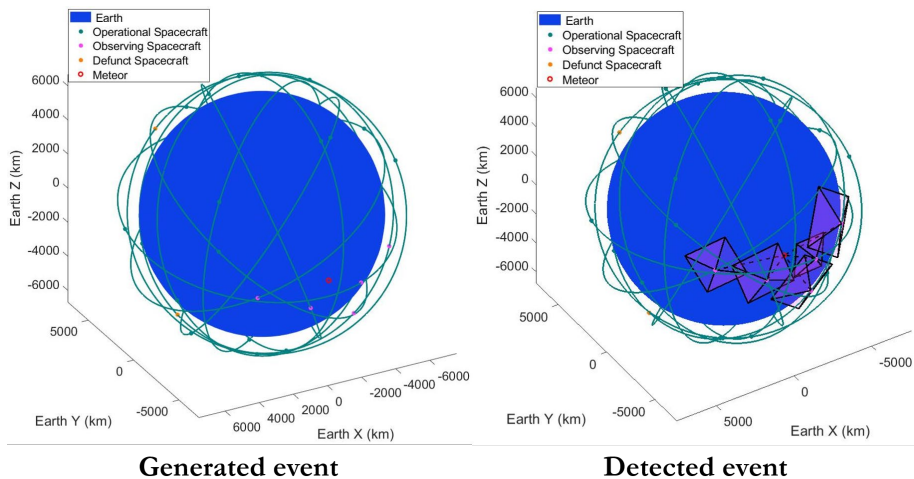

Fig. 9. The performance of the selected optimal constellation: a randomly created meteor event (left) falls in the FoV of at least three spacecraft (right).

The chain access from the constellation to the meteors was computed to verify if the meteors were accessible by at least three spacecraft. The results indicated that most random simulated meteor events were able to be successfully detected by at least three spacecraft throughout their entire flight path. The result of three random meteor event detections, along with the optimal constellation which is visualized in STK is presented in Fig. 10.

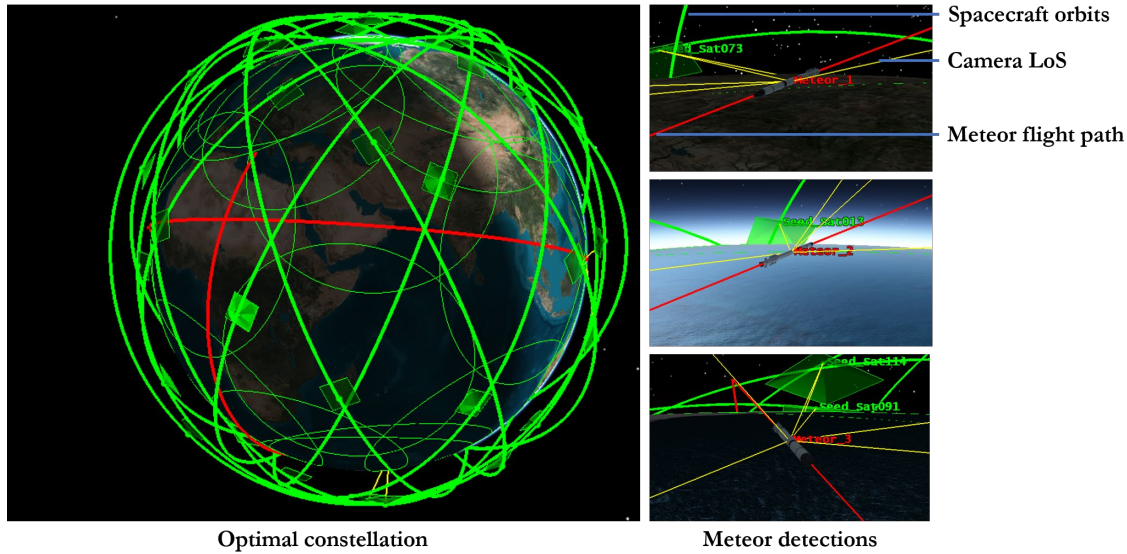

Optimal constellation    Meteor detections

Fig. 10. A visualization of the optimal meteor detection constellation (left) along with three random meteor entry detection events (right).

## 5. CONCLUSION

In this work, we developed novel algorithms and simulation architectures for constellations intended to detect meteor entry events. We began by proposing that a space-based visual meteor monitoring network is more effective than ground-based networks due to limited FoV of ground observers and atmospheric interferences. We then derived the meteor localization algorithm when the event was observed by two or more spacecraft. The derivation was based on a similar algorithm used by state-of-the-art ground-based meteor detection networks but is different due to the availability and usage of information such as the location of the observer, and the obtained information. Then a Monte-Carlo simulation architecture was developed to study the effectiveness of the algorithm. The simulation provides a sensitivity analysis of the event localization error to parameters such as number of observing spacecraft, and their pointing accuracy in measuring the direction of the meteor. The results of this study indicate that the accuracy starts to improve when at least three spacecraft are used to localize an event. For instance, if three spacecraft with pointing accuracies less than 2 deg were used, the meteor event can be localized with a maximum error of about 300 km. The localization error was shown to reduce significantly when a greater number of spacecraft were involved in observing the event. We then proceeded to design a constellation, which would guarantee that a random meteor entry event would be detected by at least three spacecraft under realistic constraints. The constellation was posed as an optimization problem which was then solved using a genetic algorithm optimizer. The results were then verified in two separate scenarios. One was using a static simulation where only detection at one time instant was checked, and the other was a dynamic simulation, where the detection was checked when the meteors have a simulated flight path, and the spacecraft in the constellation were moving at their orbital velocity. The result indicated that the proposed optimal designs were indeed able to meet their effectiveness requirement.

In this current work, we improve the state-of-the-art through four new contributions. The first is to develop the localization algorithms which can be deployed on a space-based meteor monitoring network. Following this, a Monte-Carlo simulation architecture is presented where the performance of the algorithm to different constellation parameters was developed. We then presented an automated architecture to design optimal real-world constellations using probabilistic methods and evolutionary algorithms. Finally, we developed a new automated architecture to dynamically validate the effectiveness of the optimal design using state-of-the-art spacecraft mission design software. Our future work will focus on developing accurate dynamical models of the meteors where characteristic features such as their brightness, fragmentation, and geometries are factored into the simulation. We will also focus on developing hardware testbeds which will help to identify actual limitations with the camera due to low-light conditions, blur effects and limitations of the spacecraft pointing system. These studies will help to identify a development pathway for a meteor monitoring constellation that will enhance space situational awareness.